\documentclass[11pt,a4paper]{article}
\usepackage[a4paper, total={6in, 9in}]{geometry}

\usepackage{graphicx} % Required for inserting images
\usepackage{cite}

\usepackage[dvipsnames]{xcolor}

\usepackage{caption}
\usepackage{subcaption}
\usepackage{float}
\usepackage{hyperref}
\usepackage{amsmath}
\usepackage{amssymb}
\usepackage{cleveref}
\usepackage{authblk}
\usepackage{soul}

\title{
Sensitivity to New Physics Phenomena in Anomaly Detection: A Study of Untunable Hyperparameters
}

\author[1]{Fernando Abreu de Souza\footnote{abreurocha@lip.pt}}
\author[1]{Maura Barros\footnote{maura.barros@cern.ch}}
\author[1]{Nuno F. Castro\footnote{nuno.castro@fisica.uminho.pt}}
\author[2,1]{Miguel Crispim Romão\footnote{miguel.romao@durham.ac.uk}}
\author[3,1]{Céu Neiva\footnote{ceu.neiva11@icloud.com}}
\author[4]{Rute Pedro\footnote{rute.pedro@cern.ch}}

\affil[1]{
 LIP --- Laborat\'orio de Instrumenta\c{c}\~ao e F\'isica Experimental de Part\'iculas, 
 Departamento de F\'isica,  Escola de Ciências, Universidade do Minho, 
 4701-057 Braga, Portugal
}
\affil[2]{
    Institute for Particle Physics Phenomenology,
    Durham University,
    Durham DH1 3LE,
    United Kingdom
}
\affil[3]{
    CeNTI --- Centre for Nanotechnology and Advanced Materials,
    R. Fernando Mesquita 2785, 
    4760-034 Vila Nova de Famalicão, Portugal
}
\affil[4]{
 LIP --- Laborat\'orio de Instrumenta\c{c}\~ao e F\'isica Experimental de Part\'iculas, 
 Avenida Professor Gama Pinto 2, 1649-003 Lisboa, Portugal
}

\begin{document}

\date{May 19, 2025}

\maketitle

\begin{abstract}
The search for physics beyond the Standard Model (BSM) at collider experiments requires model-independent strategies to avoid missing possible discoveries of unexpected signals. Anomaly detection (AD) techniques offer a promising approach by identifying deviations from the Standard Model (SM) and have been extensively studied. The sensitivity of these methods to untunable hyperparameters has not been systematically compared, however. This study addresses it by investigating four semi-supervised AD methods -- Auto-Encoders, Deep Support Vector Data Description, Histogram-based Outlier Score, and Isolation Forest -- trained on simulated SM background events. In this paper, we study the sensitivity of these methods to BSM benchmark signals as a function of these untunable hyperparameters. Such a study is complemented by a proposal of a non-parametric permutation test using signal-agnostic statistics, which can provide a robust statistical assessment.
\end{abstract}

%%%%%%%%%%%%%%%%%%%%%%%%%%%%%%%%%%%%%%%%%%%%%%%%%%%%%%%%%%%%%%%%%%%%%%
\section{Introduction}
\label{sec:intro}

The Standard Model (SM) of particle physics has been remarkably successful in describing a wide range of experimental results. However, it fails to account for several observed phenomena~\cite{bsm_2012}, motivating ongoing searches for clues beyond the Standard Model (BSM). At the LHC, a broad program of BSM searches is conducted, typically based on specific signal models and event topologies. Ideally, however, these searches should be as general as possible to avoid missing possible discoveries of unexpected signals.

To address this, several efforts have focused on model-independent search strategies~\cite{d0modelind1_2001,d0modelind2_2001,cdfmodelind1_2008,cdfmodelind2_2009,heramodelind1_2004,heramodelind2_2009,atlasmodelind_2019,cmsmodelind_2021}. Nevertheless, such approaches do not guarantee sensitivity to all possible BSM scenarios. Ensuring that search strategies remain broadly sensitive is therefore essential. One promising approach is the use of anomaly detection (AD) techniques, which aim to identify deviations from the expected distributions in datasets assumed to consist mostly of ``normal" (i.e., SM-like) events. Community efforts such as the LHC Olympics 2020 Anomaly Detection Challenge~\cite{lhc_olympics_2021} and the Dark Machines collaboration~\cite{darkmachines_2022} have explored a wide range of unsupervised and weakly supervised machine learning techniques for AD in collider data. These techniques have since been adopted by the ATLAS and CMS collaborations for applications ranging from BSM searches~\cite{atlas_ad_2023,atlas_unsuper_2024,atlas_weakly_2025} to jet substructure~\cite{cms_jet_2024} and data quality monitoring~\cite{cms_dq_2024}.

One of the early proposals in this direction by some of the authors, presented in~\cite{Crispim_Rom_o_2021}, used shallow and deep AD methods to search for new phenomena and compare the obtained limits with a dedicated supervised deep neural network (DNN). These models have been used to enhance the sensitivity to new physics signals~\cite{Farina_2020,anomaly_awareness_2025,adfilter_2025}, study relations between different features~\cite{CrispimRomao:2023ssj}, identify jets and anomalies within jet substructure~\cite{Heimel_2019,roy2020robust,Finke_2021,quenchedjets_2021,Canelli_2022}, triggering events~\cite{Govorkova_2022,normae_trigger_2023,hl_lhc_trigger_2023}, and modelling and simulation~\cite{modeling_2022,fastjet_2022}.

While AutoEncoders (AEs) remain the most commonly used AD method, others have been explored. Histogram-based Outlier Scores (HBOS) have been used as a novelty detector to improve the exploration capabilities of artificial intelligence-guided BSM parameter space scans~\cite{Romao:2024gjx,deSouza:2025uxb,deSouza:2025bpl}. Deep Support Vector Data Description (Deep-SVDD) has also been adapted for collider data~\cite{caron_dsvdd_2025}, where a supervised classifier is reformulated into an unsupervised anomaly detector. Weakly supervised anomaly detection methods have been applied to BSM searches as well~\cite{weaklysupervised_prior_2025,weaklysupervisedad_2025}. In parallel, recent studies have investigated contrastive learning~\cite{contrastive_neural_2025}, topology-aware architectures~\cite{topology_2025}, and symmetry-informed representations~\cite{symmetriesdatasets_2025} to improve the structure and interpret the latent space. Despite these developments, a systematic comparison of different AD models across diverse signal scenarios and key untunable parameters remains missing. 

At the same time, statistical interpretability remains a key challenge in model-independent searches. Methods that rely on anomaly scores frequently lack well-defined significance measures, making it difficult to assess discovery potential or compare across models. Recent efforts have focused on improving the statistical interpretation of AD results in the context of signal-agnostic searches~\cite{signalagnostictest_2024,goodnessfit_2024}. 

This work presents a comparative study of four anomaly detection methods -- AE, Deep-SVDD, HBOS, and Isolation Forest (iForest) -- and investigates their performance as a function of key untunable parameters. To complement performance metrics such as receiver operating characteristic (ROC) curves, we explore the M$\Delta$ and Cramér’s statistics, which allow to compare two sample distributions in a signal-agnostic way. Moreover, these are employed in a statistical permutation test for hypothesis testing, evaluating the sensitivity of each semi-supervised method studied. This non-parametric test provides a robust way to quantify deviations from the SM without relying on signal-specific assumptions, reinforcing the generality of our methodology.

The paper is organised as follows. In~\cref{sec:dataset}, we introduce the dataset of simulated events, comprising six different benchmark BSM signals and a SM background, all sharing a common final state with two leptons, one bottom jet, and large $H_T$. In~\cref{sec:methods}, we describe the semi-supervised AD methods employed in this study, including both shallow and deep approaches. These methods are used to construct discriminants for generic new physics searches. In~\cref{sec:hyper}, we assess the sensitivity of each method to different BSM signals as a function of key untunable hyperparameters, using the area under the ROC (ROC AUC) metric. In~\cref{sec:test}, we introduce a statistical framework using permutation tests to evaluate the significance of observed deviations. Finally, in~\cref{sec:conclusions}, we conclude.

\section{Dataset Simulation Details}
\label{sec:dataset}

The dataset used consists of simulated proton-proton collisions at a centre of mass energy of $13$~TeV. Events were generated at leading order using \textsc{MadGraph5}~\cite{madgraph}, with \textsc{Pythia 8}~\cite{pythia} to simulate the parton shower and hadronisation. The detection of the collision products was accomplished with the \textsc{Delphes 3}~\cite{delphes} parametrised response using the default configuration, matching the CMS detector parameters. Jets and large-radius jets are reconstructed using the anti-$\kappa_t$ algorithm~\cite{Cacciari:2008gp} with a radius parameter of $R =0.5$ and 0.8, respectively.

A diverse set of BSM signals was simulated to benchmark the anomaly detection sensitivity study in different scenarios, from new resonances to new interactions only inducing small deviations from SM predictions. These signals, used solely for evaluation and not for defining or training the methods, include:

\begin{itemize}
\item Heavy vector-like quarks (HQ): pair production of heavy vector-like $T$ quarks, with $T$ masses $m_T = \{1.0, 1.4\}$~TeV~\cite{VLQcombo};
\item Flavour changing neutral current (FCNC): $tZ$ production through a FCNC vertex~\cite{FCNCmodel2};
\item Randall-Sundrum (RS): production of an RS radion $R$ that decays into a pair of $Z$ bosons and has a mass of $m_R = 4$~TeV~\cite{Randall_Sundrum};
\item Two-Higgs-doublet model (2HDM): top quark pair production in association with a heavy Higgs boson $H'$ in the 2HDM framework, with $m_{H'} = 400$~GeV, inspired by the signal used in~\cite{2hdm};
\item Left-Right Symmetric Model (LRSM): production of the right-handed counterpart of the $W$ boson, $W_R$, decaying into a right-handed heavy neutrino, $N_R$, and a charged lepton, with masses $m_{W_R} = 6.5$~TeV and $m_{N_R} = 1.5$~TeV, inspired on the signal used in~\cite{lrsm}.
\end{itemize}

The chosen signals have a common final state consisting of $2$ leptons, $1$ bottom jet and large $H_T$ ($>500$~GeV).\footnote{$H_T$ is the scalar sum of transverse momentum ($p_T$) of all reconstructed particles in the event.} This motivates the definition of a broad phase space to focus our AD study. The SM processes constituting the background to these signals are $Z+$jets, top pair, and diboson production. To ensure a very good statistical representation, these processes were generated in samples of the kinematic space using event generation filters at parton level, as detailed in~\cite{prev_zenodo}. 

In total, $14.6$~M events were simulated, split per process type as shown in \cref{tab:events}. The background processes composing the SM cocktail, used to train the AD methods, are normalised to the expected yield using the generation cross-section for each process computed at leading order with \textsc{MadGraph5} and assuming a target luminosity of 150~fb$^{-1}$.

Each event is described by $32$ features comprising the 4-momenta, in Cartesian basis, $(p_x,p_y,p_z,m)$ in order to remove ambiguities related to periodic variables, such as $\phi=\{0,2\pi\}$\footnote{In collider physics experiments, the coordinate $\phi$ represents the azimuthal angle in the plane transverse to the colliding beam.}, which can lead to a topological obstruction in the dataset as the identification $\phi \sim \phi + 2 \pi$ is not continuous or differentiable.

The background and signal samples can be found in~\cite{new_dataset_2025}.

\begin{table}[]
    \centering
    \begin{tabular}{cc}
    \hline\hline
    Sample & Simulated Events\\
    \hline
    Background & 12.5~M \\
    $\mathrm{HQ_{1.0 TeV}}$ & 500~k\\
    $\mathrm{HQ_{1.4 TeV}}$ & 500~k\\
    FCNC & 500~k\\
    RS   & 150~k\\
    2HDM & 300~k\\
    LRSM & 150~k\\\hline\hline
\end{tabular}
    \caption{Total number of simulated events for the background sample and for each of the signal models.}
    \label{tab:events}
\end{table}

%%%%%%%%%%%%%%%%%%%%%%%%%%%%%%%%%%%%%%%%%%%%%%%%%%%%%%%%%%%%%%%%%%%%%%
\section{Semi-supervised Anomaly Detection: Models and Methodology}
\label{sec:methods}

\subsection{Semi-supervised learning methods}
\label{subsec:semisuper}

We use different shallow and deep learning algorithms to evaluate the sensitivity of semi-supervised learning methods to new phenomena. The different algorithms were only trained on simulated SM events, i.e. the background, and their sensitivity to identify examples of potential new physics events was assessed using the previously mentioned benchmark signals. These were not used during training. Two shallow and two deep learning approaches are considered in this study, which were already discussed in~\cite{Crispim_Rom_o_2021}. The shallow methods considered, HBOS~\cite{hbos} and iForest~\cite{iforest}, detect anomalies based on density or isolation principles, often requiring less computational power. The deep learning methods, Deep-SVDD~\cite{dsvdd} and AE, leverage DNN to learn more complex feature representation but, usually, require more computational power and training times.

\vspace{5mm}
\noindent \textbf{Auto-Encoder}
\vspace{2mm}

\noindent An AE is a DNN trained to reconstruct data that is compressed in a bottleneck layer. It consists of an encoder, which compresses the input data into a lower-dimensional representation, the latent space bottleneck, and a decoder, which reconstructs the original data from this latent representation. The training process is done so that the reconstruction error is minimised, according to the loss function
\begin{equation}
     L = \mathbb{E}_{\mathbf{x}\sim{\text{SM}}}[ ||\text{AE}(\mathbf{x}, \mathcal{W}) - \mathbf{x}||^2 ]
    \label{eq:mse}
\end{equation}
where $\mathcal{W}$ are the learnable parameters of the AE, $\mathbf{x}$ is the feature vector of a SM event, and $\mathbb{E}_{\mathbf{x}\sim\text{SM}}$ denotes the expected value over SM events, which in this work is obtained through a weighted average over the training set. The reconstruction error quantifies the difference between the decoded and original data and it can be used as an anomaly score, as the AE is expected to learn the relations between different features from SM processes which might not hold to new physics processes.

\vspace{5mm}
\noindent \textbf{Deep Support Vector Data Description}
\vspace{2mm}

\noindent The Deep-SVDD also uses a DNN for the training. It maps the data into a hypersphere around its centre of mass in a latent space, identifying anomalies as points lying far from the centre. The training is done to minimise the distance of all points of the training set to this centre, as expressed by the loss
\begin{equation}
    L = \mathbb{E}_{\mathbf{x}\sim{\text{SM}}}[||\text{Deep-SVDD}(\mathbf{x}, \mathcal{W}) - \textbf{c}||^2]
\end{equation}
where $\mathcal{W}$ are the Deep-SVDD trainable weights, $\textbf{c}$ is the centre of mass distribution in the output space, $\mathbf{x}$ is the feature vector of a SM event, and $\mathbb{E}_{\mathbf{x}\sim\text{SM}}$ denotes the expected value over SM events, which in this work is obtained through a weighted average over the training set. $\textbf{c}$ is obtained by passing the whole dataset through the Deep-SVDD after initialisation, but before any training takes place, and then taking the (weighted) average over all the embedding vectors, producing a centre of mass of the training data in the target latent space. The loss minimisation forces the network to find common patterns in data as to successfully embed it into the target space. The anomaly score in a Deep-SVDD is how far the event is from the centre $\textbf{c}$.

\vspace{5mm}
\noindent \textbf{Isolation Forest}
\vspace{2mm}

\noindent The iForest algorithm is a tree-based AD method that isolates anomalies by successive random partitions of sub-samples of the data. The partitions are represented as trees, which are grown until a maximum depth is achieved, or no further splitting is possible. As the partitions are random, the number of nodes of the tree that a data example needs to traverse until it reaches a leaf node is a measurement of how inlier it is. Conversely, outliers require fewer splits to be isolated, arriving at a leaf node after traversing fewer nodes. The discriminant is therefore the iForest score, which takes the form
\begin{equation}
    \text{iForest}(\mathbf{x}) =  2^{-\frac{\mathbb{E}[h(\mathbf{x})]}{d}} \ , 
\end{equation}
where $d$ is the average traversable path in a binary tree of the same depth, $h(\mathbf{x})$ is the number of nodes that a data example with a feature vector, $\mathbf{x}$, has to travel in a given tree, and $\mathbb{E}$ is the average over all the trees in the forest~\cite{iforest}. iForests were recently proposed as an anomaly detection model in the context of transient detection in microlensing data~\cite{CrispimRomao:2025pyl}.

\vspace{5mm}
\noindent \textbf{Histogram-based Outlier Score}
\vspace{2mm}

\noindent HBOS is perhaps the most simple of all the semi-supervised models considered in this work. It computes a histogram for each feature while assuming feature independence. Unlike methods based on pairwise comparison or distance measure, HBOS estimates the probability density per histogram bin. To compute the anomaly score, the probability of each feature value falling into its respective bin is determined and a score of $\log_2(\text{HIST})$ is associated, where HIST is the density (height) of that bin. The total anomaly score is given by the sum across all features. 

\subsection{Training and Tunable Hyperparameters}

The dataset was split equally into train, validation and test sets to guarantee statistical representativeness at each stage. Only SM events were used to train the semi-supervised methods. To account for statistical fluctuations of the data and the stochastic character of the semi-supervised methods, a collection of 10 independent models was trained for each method, each training using a different subset of the training dataset and with a different model initialisation, when applicable. The hyperparameters of some of the models were optimised, using performance metrics evaluated on the validation set. The final model evaluation, analysis, and statistical test were performed on the test set. All features were standardised setting their mean to zero and standard deviation to unity.

The choice of a model's hyperparameters can have a huge impact on its performance, making hyperparameter tuning a crucial step in optimising the model's performance. For supervised tasks, the hyperparameters can be efficiently tuned by using techniques like grid search or Bayesian optimisation. However, for semi-supervised tasks, one might not have a sensible and meaningful metric to validate a choice of hyperparameters. In the work at hand, this is evident, as certain parameters are more difficult to fine-tune: the latent space size of AE and Deep-SVDD, the number of bins of HBOS, and the number of estimators in iForest. As our methodology is strictly semi-supervised, where the AD models are only trained on SM processes, we lack a clear validation criterion on how well an AD model separates SM from BSM events. Consequently, these hyperparameters are typically selected at the beginning of the analysis, before further model optimisation, based on the dataset properties and dimensionality. A key goal of this work is to provide a comparison of the sensitivity of the different methods to new phenomena for different values of these so-called untunable hyperparameters. We will focus on the values of these untunable hyperparameters presented in~\cref{tab:untunable}, and the results are presented in~\cref{sec:hyper}.

The HBOS method was based on the \texttt{pyod} Python toolkit~\cite{pyod}, with sample normalisation weights incorporated when computing the histograms. The iForest implementation was based on \texttt{Scikit-Learn}~\cite{sklearn}. For both shallow methods, a principal component analysis (PCA) was applied to remove the linear correlations between the features.\footnote{This is especially important for HBOS as it assumes feature independence. For the iForest, we found in the early stages of this work that it helps boost discrimination. The reason for this is that the partitions are along the feature axis, which can lead to an excessive coverage of the feature space if there are strong linear correlations between features. By removing the linear dependency of the features, the iForest will partition along the principal components of the dataset, increasing the sensitivity outside of these directions, which in turn makes the iForest more sensitive to new phenomena where correlations between features might differ.} The PCA was implemented using \texttt{Scikit-Learn}. The values considered are represented in~\cref{tab:untunable}.

\begin{table}[h!]   
    \begin{center}
        \begin{tabular}{ c  c  c}
        \hline\hline
            Model   & Untunable Hyperparameter & Considered Values \\
            \hline
            HBOS    & Number of Bins           & $[1, 31]$ \\
            iForest & Number of Estimators     & $[25, 200]$ in steps of $25$ \\
            AE      & Latent Space Dimension   & $[1, 31]$ \\
            Deep-SVDD   & Latent Space Dimension   & $[1, 31]$ \\ \hline\hline
        \end{tabular}
    \end{center}
    \caption{The untunable parameters of the different AD models.}
    \label{tab:untunable}
\end{table}

The deep models were implemented with \texttt{TensorFlow 2.11}~\cite{tf}. Several were trained to scan the AE and the Deep-SVDD latent space dimension across the range defined in~\cref{tab:untunable}. For each value of the untunable hyperparameters, \texttt{optuna}~\cite{optuna} was used to optimise the tunable hyperparameters, within the search space detailed in~\cref{tab:hyper_ae} and~\cref{tab:hyper_dsvdd}.

The AE was optimised over $1000$ epochs, with early stopping applied using patience of $50$ epochs, and $100$ \texttt{optuna} trials to maximise the quality of the reconstruction as measured by the coefficient of determination\footnote{With $R^2 = 1 - \mathbb{E}[||x-\hat x||^2]/\mathbb{E}[||x-\mathbb{E}[x]||^2]$, where $x$ are the feature vectors and $\hat x$ their reconstruction, with the expected value being computed using the weights. Intuitively, the coefficient of determination quantifies the ratio of variance that is being described by the regression.}, $R^2$, as the training process can be seen as a supervised regression problem. Therefore, for a fixed dimension of the latent space, there is a meaningful metric to validate and compare different values of the remaining hyperparameters in~\cref{tab:hyper_ae}. Furthermore, to prevent features with higher nominal values from dominating the loss function, the AE was trained to reconstruct the standardised version of the features.

\begin{table}[h!]   
    \begin{center}
        \begin{tabular}{ c  c }
            \hline\hline
            Hyperparameter & Possible Values \\
            \hline
            Number of Layers & [1, 10] \\
            Number of Units & [8, 512] in steps of 8 \\
            Dropout Rate & [0, 0.5] in steps of 0.1 \\
            Activation Function & ReLU or LeakyReLU \\
            Normalisation & BatchNormalization, LayerNormalization or None\\\hline\hline
        \end{tabular}
    \end{center}
    \caption{Hyperparmeter search space for the AE.}
    \label{tab:hyper_ae}
\end{table}

For the Deep-SVDD, for each value of the latent space dimension we optimised the rest of the hyperparameters in~\cref{tab:hyper_dsvdd} over $10~000$ epochs, with early stopping applied using patience of $100$ epochs, and $100$ trials to minimise the average distance to the centre. The best hyperparameter combination resulting from the optimisation loop for each value of the untunable parameter was selected and used to train 10 models, each on an independent subsample of the training set.
This allows us to estimate the statistical uncertainty as the standard deviation of the results obtained from the model collection.

\begin{table}[h!]   
    \begin{center}
        \begin{tabular}{ c  c }
        \hline\hline
            Hyperparameter & Possible Values \\
            \hline
            Number of Layers & [1, 10] \\
            Number of Units & [8, 512] in steps of 8 \\
            Dropout Rate & 0 \\
            Activation Function & ReLU or LeakyReLU \\
            $\beta_1$ & [0.85, 0.95] \\
            $\beta_2$ & [0.99, 0.9999] \\
            Weight Decay & 0.0, 1e-9, 1e-8, 1e-7, 1e-6, 1e-5, 1e-4 or 1e-3\\\hline\hline

        \end{tabular}
    \end{center}
    \caption{Hyperparmeter search space for the Deep-SVDD.}
    \label{tab:hyper_dsvdd}
\end{table}

%%%%%%%%%%%%%%%%%%%%%%%%%%%%%%%%%%%%%%%%%%%%%%%%%%%%%%%%%%%%%%%%%%%%%%%%

\section{Sensitivity dependency on the Untunable Hyperparameter}
\label{sec:hyper}

In this section, we evaluate how the choice of untunable hyperparameters, presented in~\cref{tab:untunable}, influences the sensitivity of each semi-supervised method to new physics phenomena using benchmark signals. To quantify sensitivity, we employ the ROC AUC, a widely used and interpretable metric in similar analyses. However, it is important to note that ROC AUC requires true labels, making it a supervised discrimination metric. In~\cref{subsec:stat}, we will explore alternative metrics suitable for signal-agnostic studies and their application in developing a semi-supervised statistical test for detecting new physics phenomena in a sample.

\begin{figure}[H]
    \centering
    \captionsetup[subfigure]{labelformat=empty}
    \makebox[\textwidth][l]{
    \begin{subfigure}[t]{0.475\textwidth}
    \includegraphics[scale=0.15,trim={0 0 16cm 0},clip]{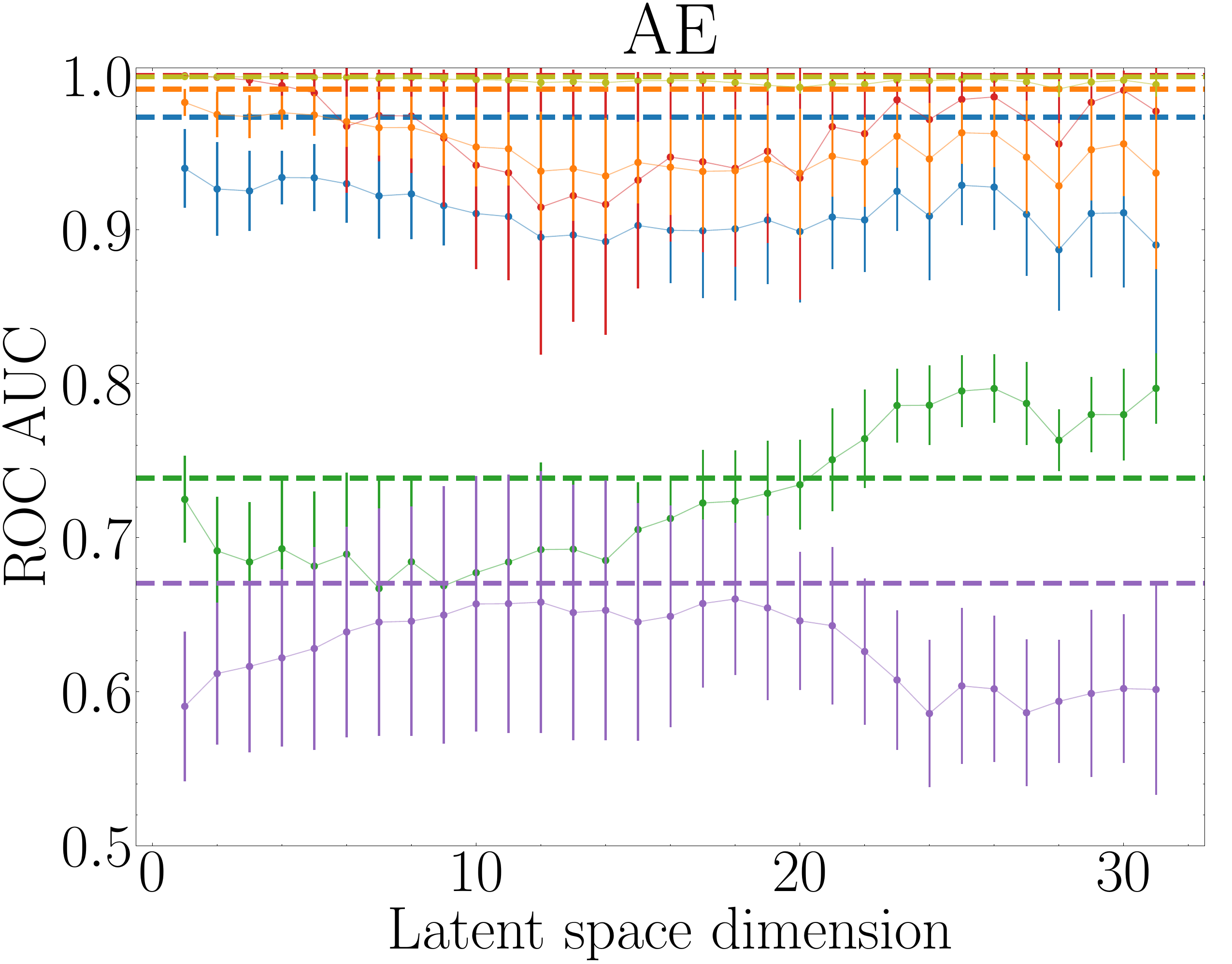} 
    \label{fig:roc_ae2} 
    \end{subfigure}
    \begin{subfigure}[t]{0.475\textwidth}
    \includegraphics[scale=0.15,trim={0 0 16cm 0},clip]{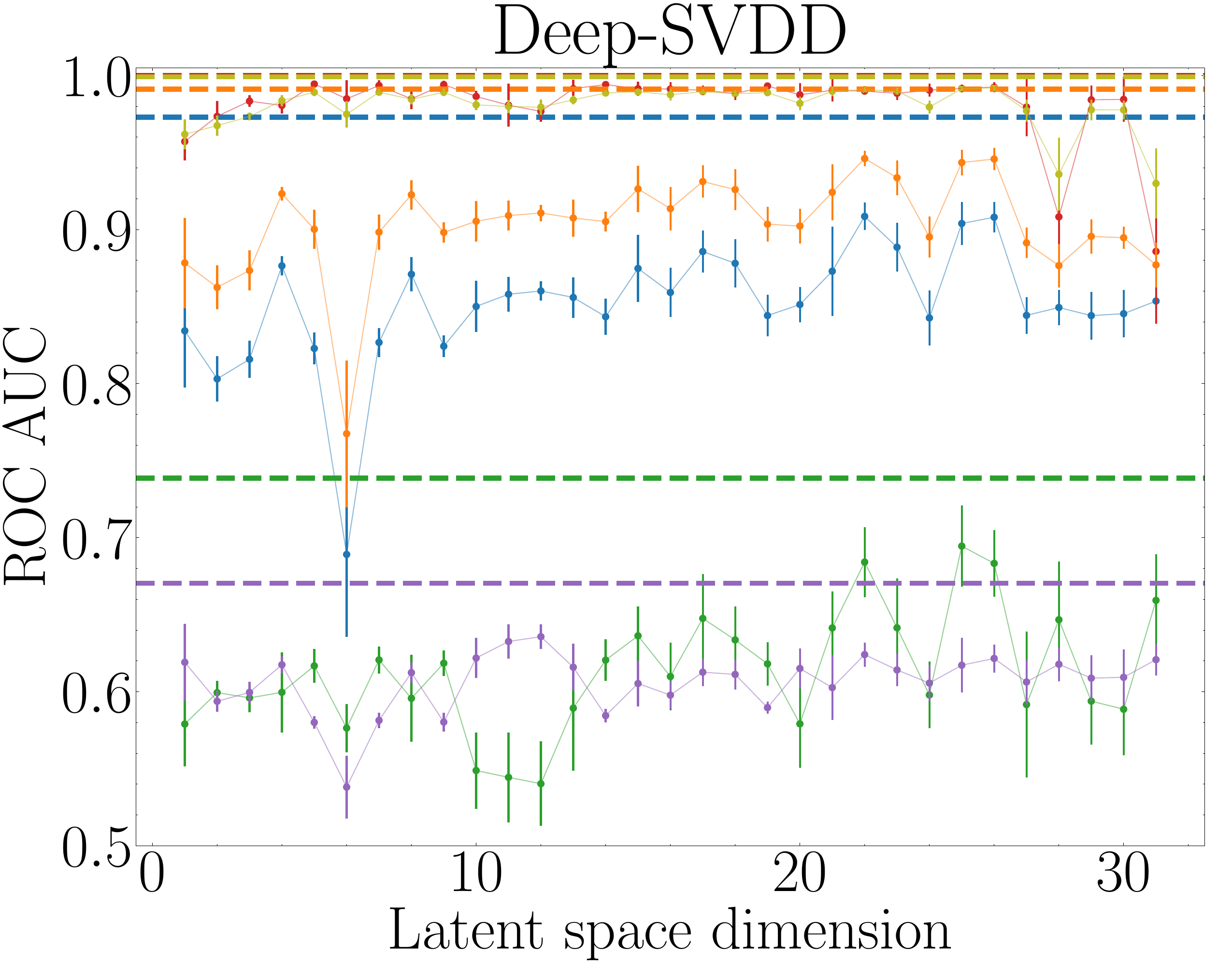} 
    \label{fig:roc_dsvdd2} 
    \end{subfigure}}
    \makebox[\textwidth][l]{
    \begin{subfigure}[t]{0.475\textwidth}
    \includegraphics[scale=0.15,trim={0 0 16cm 0},clip]{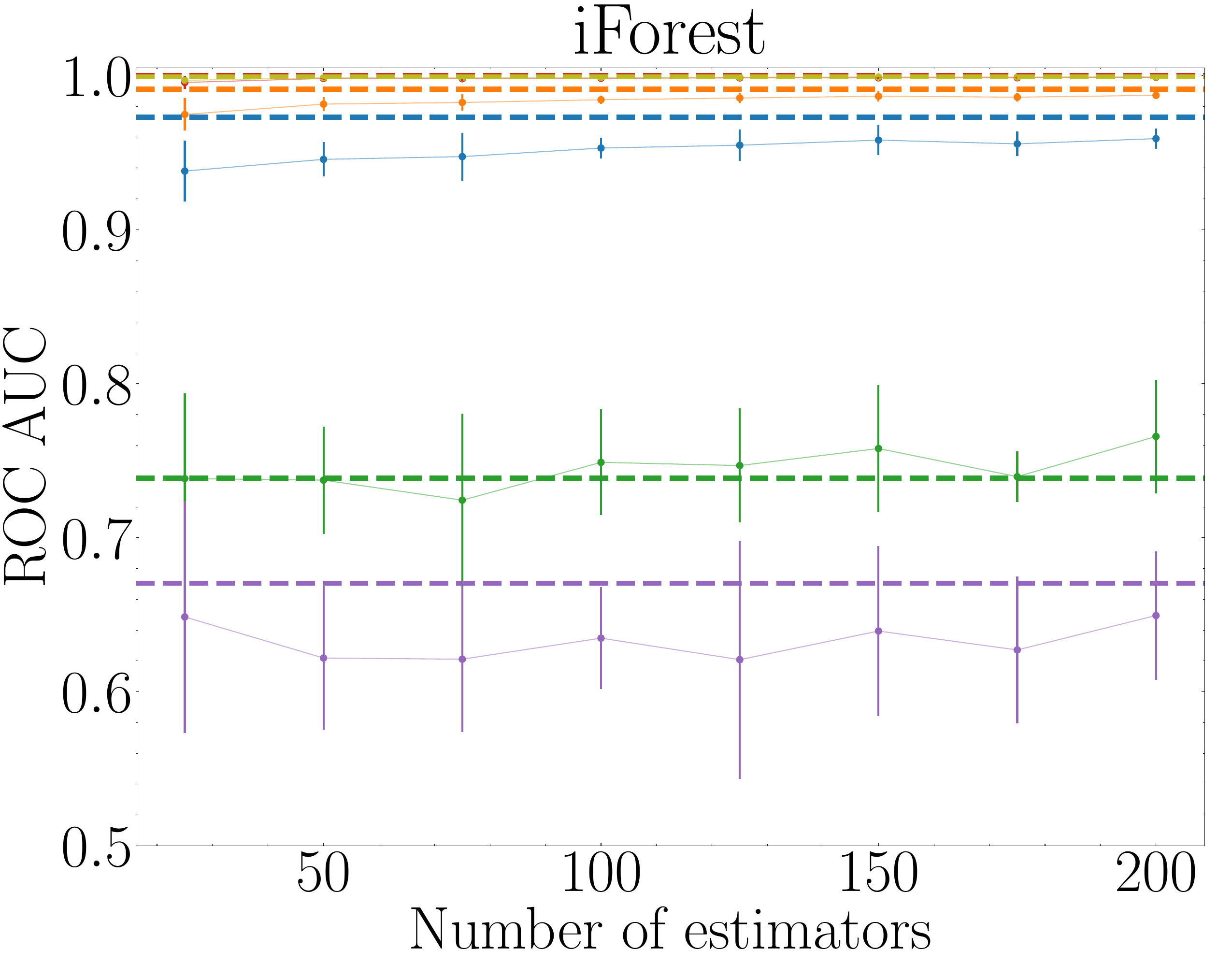} 
    \label{fig:roc_iforest2} 
    \end{subfigure}
    \begin{subfigure}[t]{0.475\textwidth}
    \includegraphics[scale=0.15]{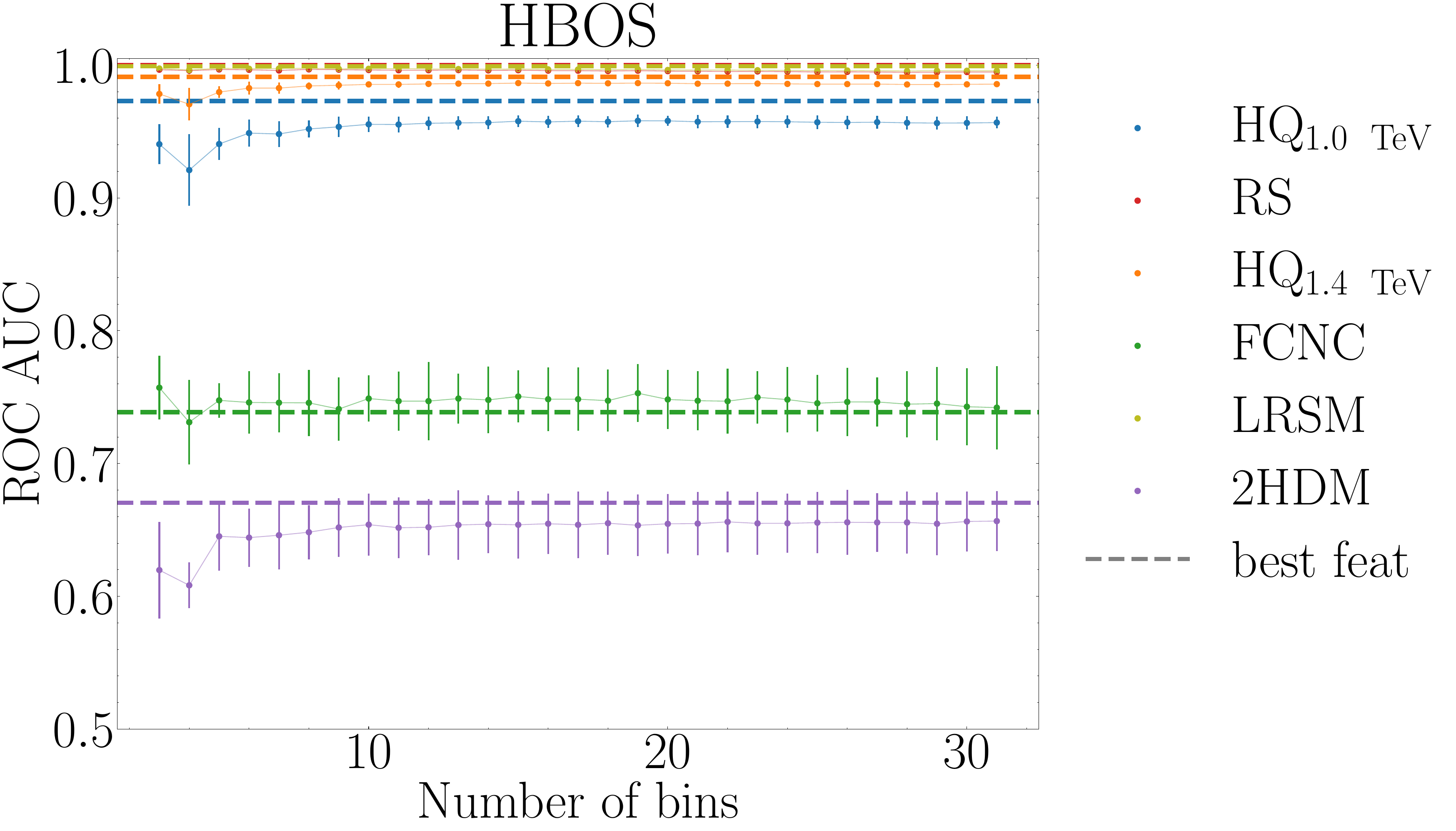} 
    \label{fig:roc_hbos2} 
    \end{subfigure}}
    \caption{ROC AUC for each signal and semi-supervised method as a function of its untunable hyperparameter. The ROC AUC for the most discriminating feature for each signal is also displayed as a dashed line. The uncertainty bars are the standard deviation of the ROC AUC values obtained from the model collection.}
    \label{fig:roc_auc}
\end{figure}

Figure~\ref{fig:roc_auc} illustrates the dependence of the ROC AUC on untunable hyperparameters for each semi-supervised method. The results indicate that deep methods (AE and Deep-SVDD) exhibit some variation in background-signal separation concerning latent space dimensionality. For instance, the AE initially shows a decrease in sensitivity for HQ signals, stabilising at a latent space dimension of approximately $\gtrsim 15$. For the FCNC (2HDM) signal, sensitivity remains nearly stable, with a slight increase (decrease) observed beyond a latent space dimension of $\gtrsim 20$. Despite these variations, the AE sensitivity changes are generally within statistical uncertainty across most signals. Likewise, the Deep-SVDD does not exhibit a monotonic relationship between ROC AUC and latent space dimensionality, maintaining a largely flat profile. This suggests that, for both deep learning models, latent space dimensionality has a minimal impact on discriminative power across a broad range of signals. Similarly, shallow methods show negligible variation, as evidenced by the stable sensitivity of HBOS and iForest to their untunable hyperparameters. In all the cases, we see that the semi-supervised methods are at best as good as the most discriminating feature for each signal, and at worst only slightly worse than the best feature. This observation is reassuring, as it means that the semi-supervised methods ``conserve'' the separation provided by the best feature without knowing which one it is as this identification is signal-dependent.

The results for AE imply that its discrimination power is largely independent of latent space dimensionality. However, this finding challenges the prevailing intuition that improved background reconstruction should enhance sensitivity to new physics phenomena. Given that the AE is trained to model non-linear relationships in SM events and detect deviations as potential indicators of new physics phenomena, one might expect a stronger correlation between reconstruction quality and sensitivity. Conversely, an AE that poorly reconstructs SM events should, at least in principle, be less effective in detecting anomalies. To investigate this, we now analyse the interplay between sensitivity and AE reconstruction quality.

\begin{figure}[H]
   \centering
   \makebox[\textwidth][c]{
   \begin{subfigure}{0.5\textwidth}
   \centering
   \includegraphics[width=1.0\textwidth]{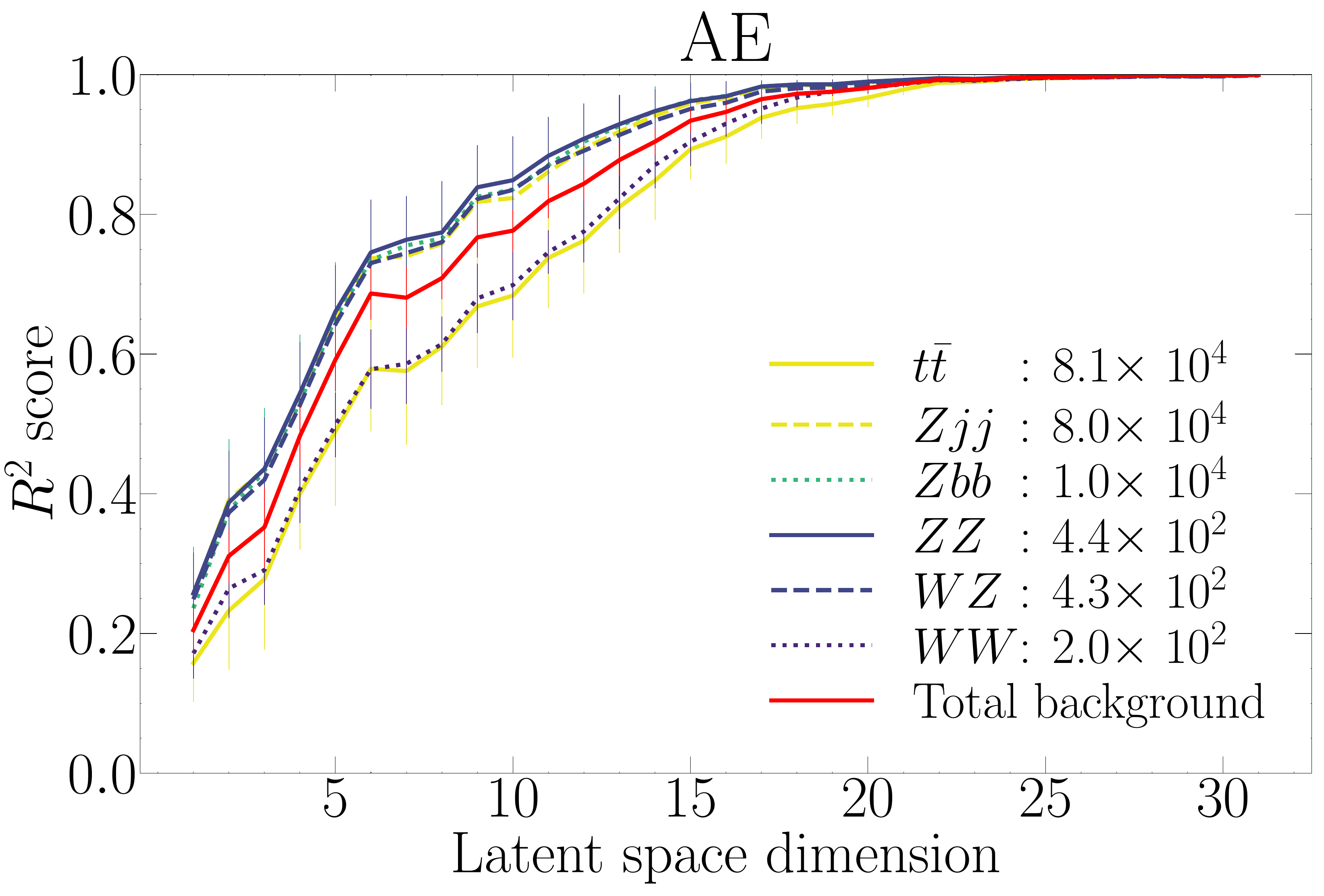}
   \end{subfigure} %\hspace{20mm}
   \begin{subfigure}{0.5\textwidth}
   \centering
   \includegraphics[width=1.0\textwidth]{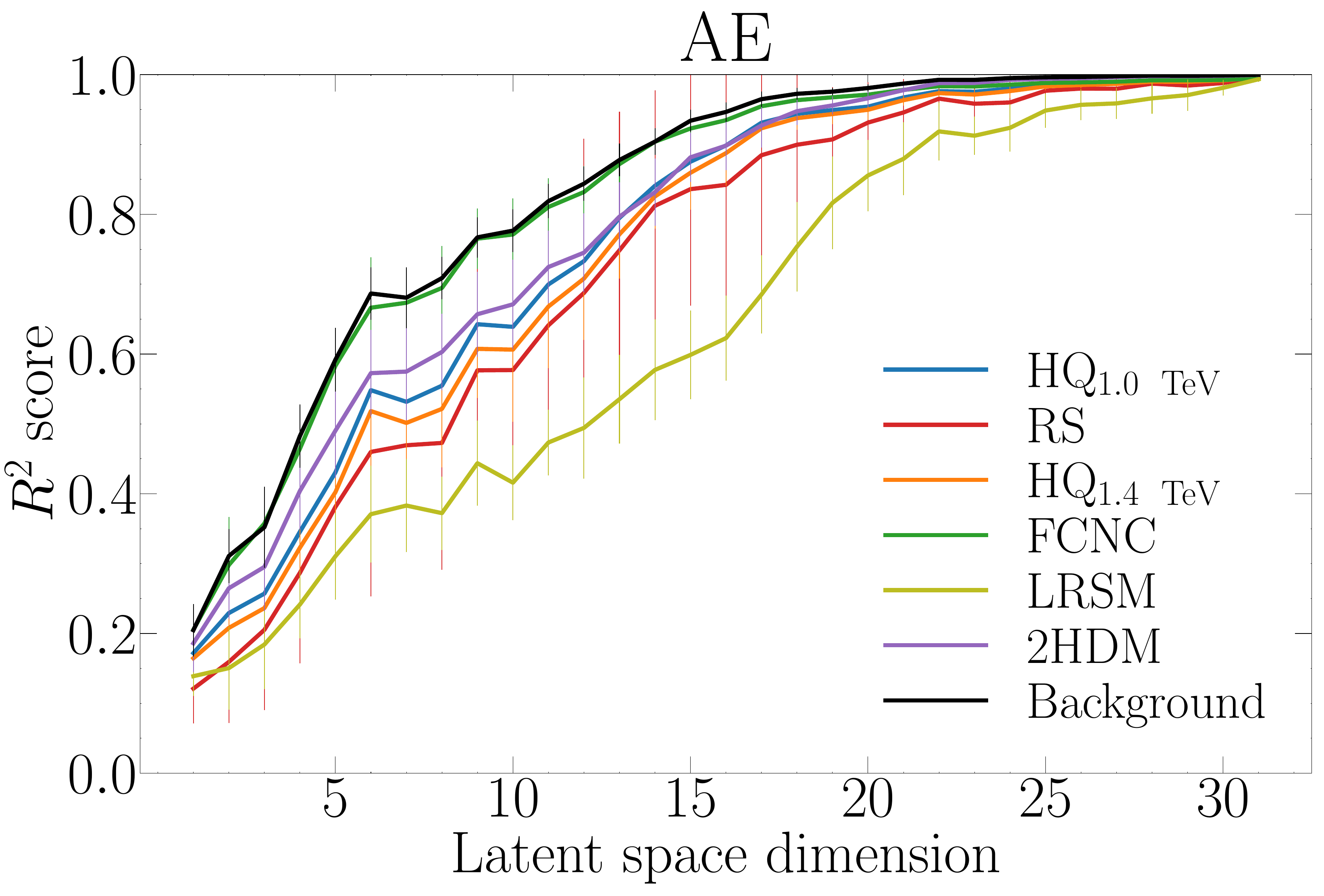}
   \end{subfigure}}
   \caption{$R^{2}$ score as a function of the latent space dimensionality of the AE. (Left) AE $R^{2}$ reconstruction score for each background process across different latent space dimensions. (Right) AE $R^{2}$ reconstruction score for each signal across different latent space dimensions. The uncertainty bars are the standard deviation of the $R^2$ values obtained from the model collection.}
   \label{fig:r2_vs_yields}
\end{figure}

In principle, the latent space dimension affects the reconstruction quality of both background and signal data. Ideally, AE models should capture the underlying physics of the SM while distinctly representing new BSM signals. However, as shown in~\cref{fig:r2_vs_yields}, a higher latent space dimension does not necessarily lead to improved background-signal separation. The reconstruction quality of the AE, quantified by the $R^{2}$ score, increases monotonically with latent space dimension for both background (left pane) and signal (right pane). This is of course expected, as a higher latent space dimensionality allows for a more complete description of the training data as less information will be lost by the encoding bottleneck. However, if higher reconstruction quality correlated with improved sensitivity, we would expect a corresponding increase in ROC AUC with latent space dimension, yet this contradicts the findings in~\cref{fig:roc_auc}. Nevertheless, we observe that, for a fixed latent space dimension, the relative reconstruction quality of different signals aligns with their ranking in terms of AE sensitivity in~\cref{fig:roc_auc}.

\begin{figure}[H]
    \centering
    \includegraphics[width=0.7\textwidth]{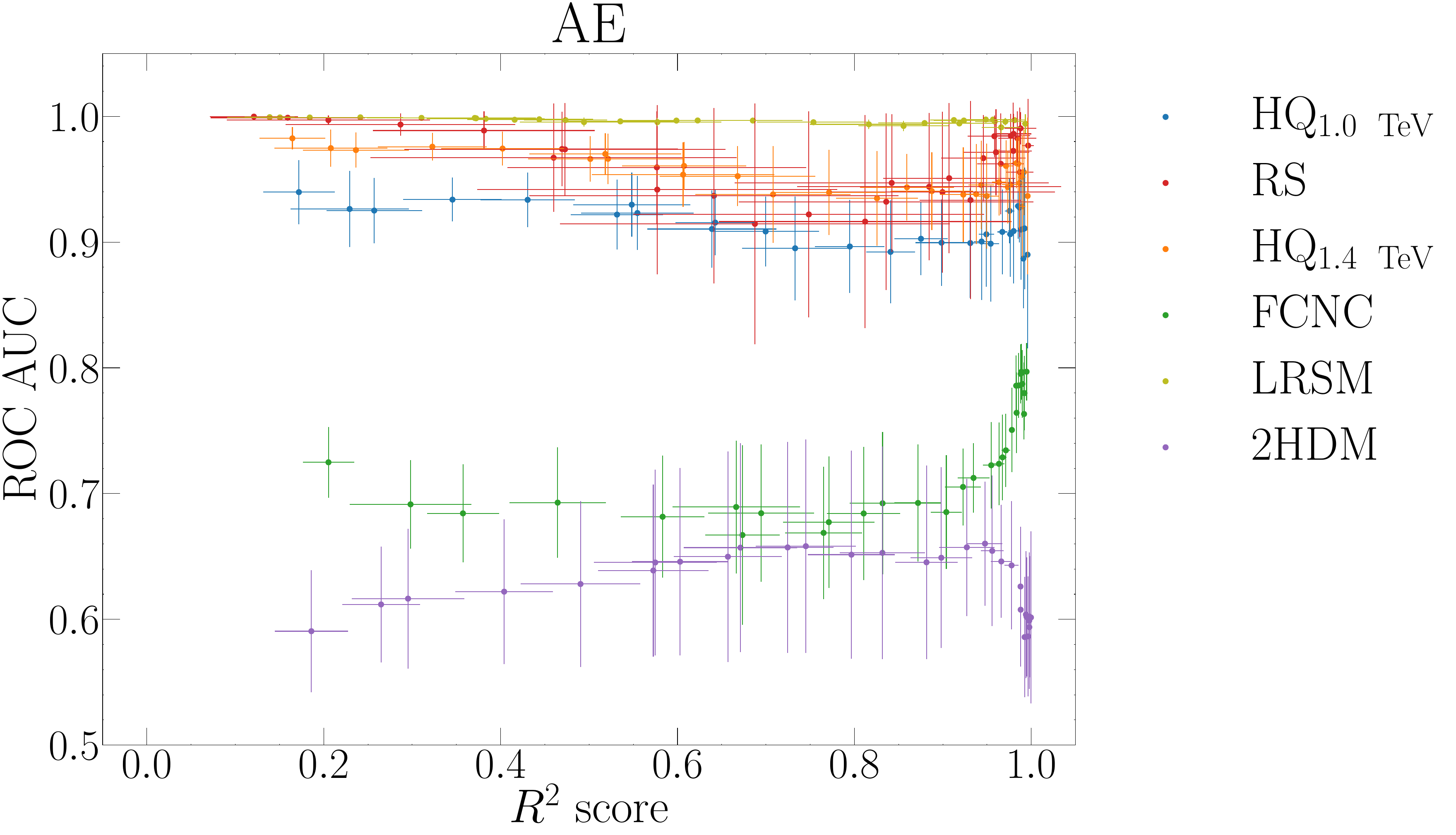}
    \caption{ROC AUC as a function of AE reconstruction score $R^2$ for each signal. The uncertainty bars are the standard deviation of the ROC AUC and $R^2$ values obtained from the model collection.}
    \label{fig:roc_vs_r2} 
\end{figure}

Figure~\ref{fig:roc_vs_r2} further supports this observation by plotting ROC AUC against the reconstructed $R^2$ score for each signal. The results reveal no monotonic correlation between ROC AUC and $R^2$. More critically, we find that discrimination power does not converge to 0.5 (random classification) as $R^2 \to 0$, nor does it approach 1.0 (perfect classification) as $R^2 \to 1$.

These observations lead to a key conclusion: while for a fixed latent space dimension, discrimination power depends on the relative difference in reconstruction error between signal and background, it does not depend on the absolute reconstruction quality of the background, which increases monotonically with the latent space dimension. This suggests that background-signal separation remains largely invariant across different latent space dimensions.

Additionally, one might assume that different background processes could have optimal AE latent space dimensions, resulting in noticeable differences in their reconstruction quality. However, the left pane of~\cref{fig:r2_vs_yields} does not support this hypothesis for the training dataset used in this analysis. Instead, reconstruction quality differences across background processes remain small up to a latent space dimension of approximately $20$, beyond which the AE achieves $R^2 \sim 1$, effectively learning how to reconstruct all background processes. This lack of variability is likely due to the stringent selection criteria applied in this analysis, which include requiring two leptons, one bottom jet, and a large $H_T$. These cuts yield highly similar events, diminishing potential differences that might otherwise lead to distinct optimal embedding dimensions.\footnote{See, for example,~\cite{topology_2025} on a discussion on the topology and geometry of the AE latent spaces for different physical processes.}

Finally, while this discussion has focused on AE, where $R^2$ provides an interpretable semi-supervised metric, the same intuition applies to Deep-SVDD. Since the Deep-SVDD latent space serves as an embedding space similar to the AE latent space, its sensitivity to new physics phenomena should also be largely invariant across different latent space dimensions, which is demonstrated in~\cref{fig:roc_auc}.

\section{Signal Agnostic Statistical Test with Permutations}
\label{sec:test}

In this section, we introduce a statistical test based on sample permutations and signal-agnostic test statistics to evaluate the sensitivity of semi-supervised models to new phenomena.  In particular, we study their dependence on untunable hyperparameters.

\subsection{Test statistics}
\label{subsec:stat}

In~\cref{sec:hyper}, we assessed the discrimination dependence of the considered semi-supervised methods to untunable hyperparameters using one of the most common metrics in machine learning classification problems, the ROC AUC. While widely used and easy to interpret, the ROC AUC is inherently a supervised metric that requires true labels, making it unsuitable as a test statistic for unexpected new phenomena. Additionally, the ROC AUC has several limitations, such as insensitivity to symmetric but distinct distributions,\footnote{Since all momentum and missing energy components are symmetric around zero, the ROC AUC can be misleading. Even when background and signal distributions differ in shape, their symmetry can cause the ROC AUC to remain close to $0.5$, failing to reflect the true discriminative power. This issue is particularly relevant when using the best feature ROC AUC as a baseline for comparison.} an implicit assumption that the class of interest has a higher median than the negative class, and a tendency to saturate quickly, especially for signals with large values in the tails of the background distribution.

To address these issues, we consider two alternative label-free test statistics: M$\Delta$ and Cramér's test (Cr), which we define below.

\paragraph{M$\Delta$:}

M$\Delta$ is inspired by the Kolmogorov-Smirnov test for comparing two sample distributions~\cite{Kolmogorov, Smirnov}. Given two samples, $A$ and $B$, M$\Delta$ for a univariate quantity $x$ is defined as:
\begin{equation}
\text{M}\Delta_{A,B} = \max_x | \text{eCDF}_{A}(x) - \text{eCDF}_{B}(x)|,
\label{eq:mdelta_def}
\end{equation}
where $\text{eCDF}_i(x),\ i=A,B$ are the empirical cumulative distribution functions (eCDFs) of $x$ in samples $A,B$. M$\Delta$ thus represents the maximum difference between two eCDFs\footnote{In this work, the eCDFs are always computed with simulation weights to reflect the true physical distributions.}. This statistic has several advantages: it is insensitive to the ordering of distributions (i.e., $\text{M}\Delta_{AB} = \text{M}\Delta_{BA}$), it can detect differences in symmetric eCDFs, and it produces values in the range $[0,1]$, making it easy to interpret and compare.

\paragraph{Cr:}

Cr was introduced in~\cite{CrD}, inspired by a one-sample goodness-of-fit test statistic proposed by Harald Cramér~\cite{Cramér}. It is defined as the integral of the squared difference between two empirical cumulative distribution functions. Given two samples, $A$ and $B$, Cr for a univariate quantity $x$ is given by:
\begin{equation}
\text{Cr}_{A,B} = \int_{-\infty}^{\infty} |\text{eCDF}_{A}(x) - \text{eCDF}_{B}(x)|^{2} dx.
\label{eq:cvm_def}
\end{equation}
%where $\text{eCDF}_i(x),\ i=A,B$ are the empirical cumulative distribution functions of $x$ in samples $A$ and $B$.
Since Cr is unbounded for unbounded $x$, it is more challenging to interpret and compare across distributions.\footnote{One possible way to improve interpretability is to scale the two sample distributions into a fixed interval. However, even with this adjustment, we found that the resulting values were not suitable for direct comparison across different discriminant samples, as the domain size varies significantly between different features and semi-supervised methods.} Nevertheless, Cr is highly sensitive to differences in symmetric distributions. Its primary advantage, compared to the other test statistics considered, is its sensitivity to the tails of the discriminant distribution -- a property that proves particularly useful when performing a two-sample permutation test, as discussed in the next section. We also notice that, although similar to the Cramér-von Mises and the 2-Wasserstein distances, the Cr test used in this work is distinct and has the benefit of being easier to compute than these other two.

\subsection{Statistical Test with Permutations}
\label{subsec:permutation_test}

The test statistics presented earlier quantify the discrimination power of semi-supervised methods at background signal separation in the same way that ROC AUC can be used in supervised methods. However, a hypothesis testing methodology is necessary to claim evidence for new phenomena in observed data.
Usually, this is derived from the ratio of likelihoods under the background-only and signal hypotheses. The question arises on how to perform such a hypothesis test in a signal-agnostic way when the signal hypothesis is unknown a priori. 

In this work, we employ the two-sample test statistics presented above with a permutation test to achieve a hypothesis testing methodology for AD High Energy Physics analyses. Permutation tests are used to assess whether two independent samples, $X$ and $Y$, obtained through some underlying distributions, $X \sim p_X$ and $Y \sim p_Y$, are in fact generated by the same distribution. Therefore, the null hypothesis for the two-sample statistical test is that the two samples come from the same underlying distribution,
\begin{equation}
    H_0: p_X = p_Y.
\end{equation}

\begin{figure}[H]
    \centering
    \includegraphics[width=0.9\textwidth]{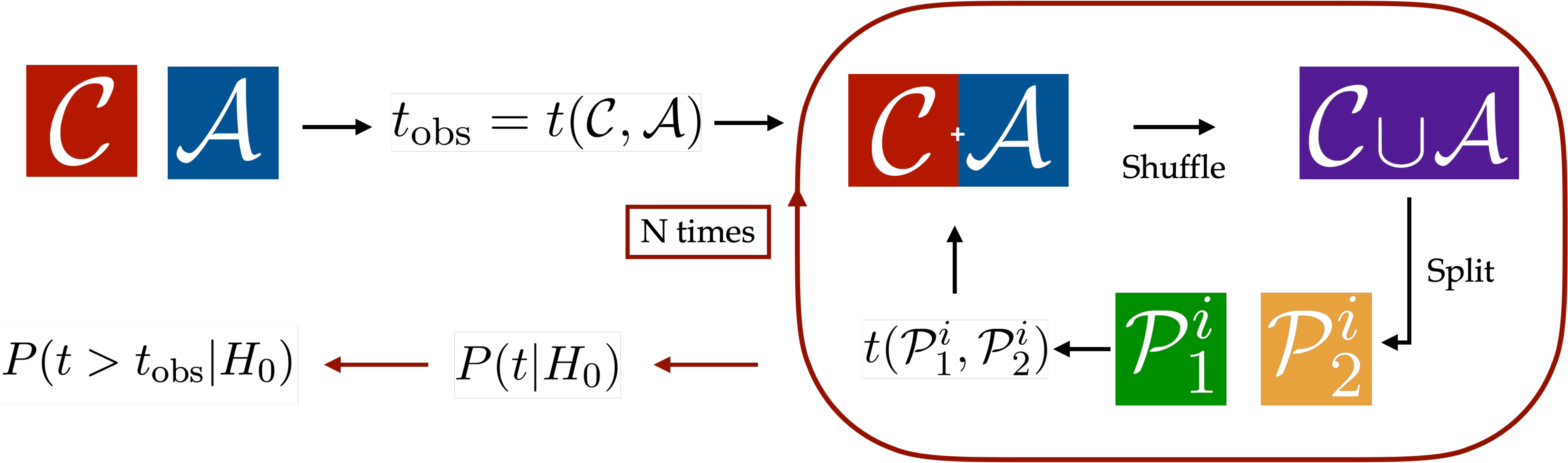}
    \caption{Statistical test with permutations.}
    \label{fig:permutation_test} 
\end{figure}

The choice of the permutation test is motivated by it being a non-parametric two-sample statistical test that can be combined with any suitable test statistic without the need for prior knowledge about the shape of distributions. This is exactly what is needed in signal-agnostic analyses where there is no knowledge about the nature of the new physics signal. The proposed methodology for the two-sample permutation test works is listed below, which is also presented in a diagrammatic form in~\cref{fig:permutation_test}:
\begin{enumerate}
    \item Consider two samples, control $\mathcal{C}$ and analysis $\mathcal{A}$, where $\mathcal{C}$ can be thought of as the background simulation produced by the experiment and $\mathcal{A}$ as the recorded experimental data;
    \item Randomly pool $\mathcal{C} \cup \mathcal{A}$ into two samples: $\mathcal{P}^i_1$, $\mathcal{P}^i_2$, where $i=1,\dots,N$ is the permutation label;
    \item Calculate $t(\mathcal{P}^i_1, \mathcal{P}^i_2)$, where $t$ is the chosen test statistic;
    \item Repeat steps $2$ and $3$ for $N$ permutations  to derive the distribution of the test statistic under the null hypothesis, $P(t|H_0)$;
    \item Compute the observed test: $t_{\text{obs}} = t(\mathcal{C}, \mathcal{A})$;
    \item The one-sided $p$-value of the test is given as $p_{\text{value}} = P(t>t_{\text{obs}}|H_0) = \frac{\#(t > t_{\text{obs}})}{N}$;\footnote{Since we are considering statistical distances for our test statistics, the evidence against the null hypothesis will be embodied by large values of the test statistic, therefore the $p$-value is computed on the right-hand side of the $P(t|H_0)$.}
    \item For $p_{\text{value}} \leq 0.05$, there is evidence to reject the null hypothesis, $H_0$, with $95\%$ confidence.
\end{enumerate}

It is important to note that, strictly speaking, the null hypothesis is whether both samples, $\mathcal{C}$ and $\mathcal{A}$, are originated by the same distribution. However, if $\mathcal{C}$ is prepared in such way that it only includes SM events, $\mathcal{C}\sim p_{SM}$, the null hypothesis becomes equivalent to the statement that data is also solely composed of SM events, i.e. $\mathcal{A}\sim p_{\text{Data}} = p_{SM}$, and therefore we obtain a semi-supervised statistical test on the presence of new phenomena in the data.\footnote{Of course, one could point out that any severe enough mismodeling can lead to a rejection of the null hypothesis, i.e. the SM. However, this is a common challenge in searches for new phenomena and not of this methodology in itself.}

\begin{figure}[h!]
    \includegraphics[width=0.48\textwidth]{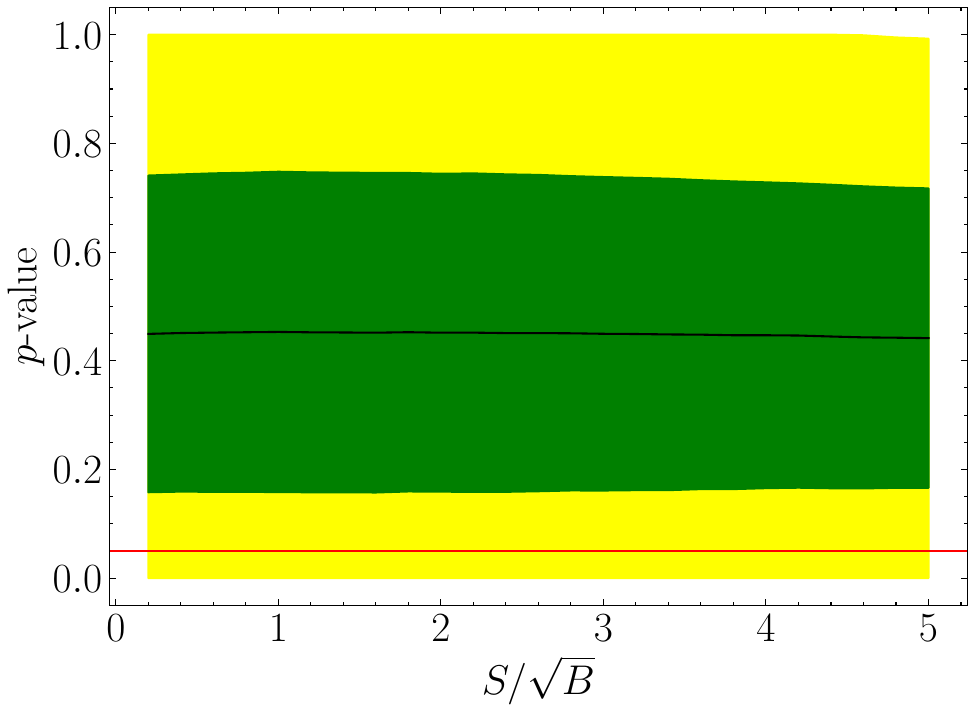}
    \hfill
    \includegraphics[width=0.48\textwidth]{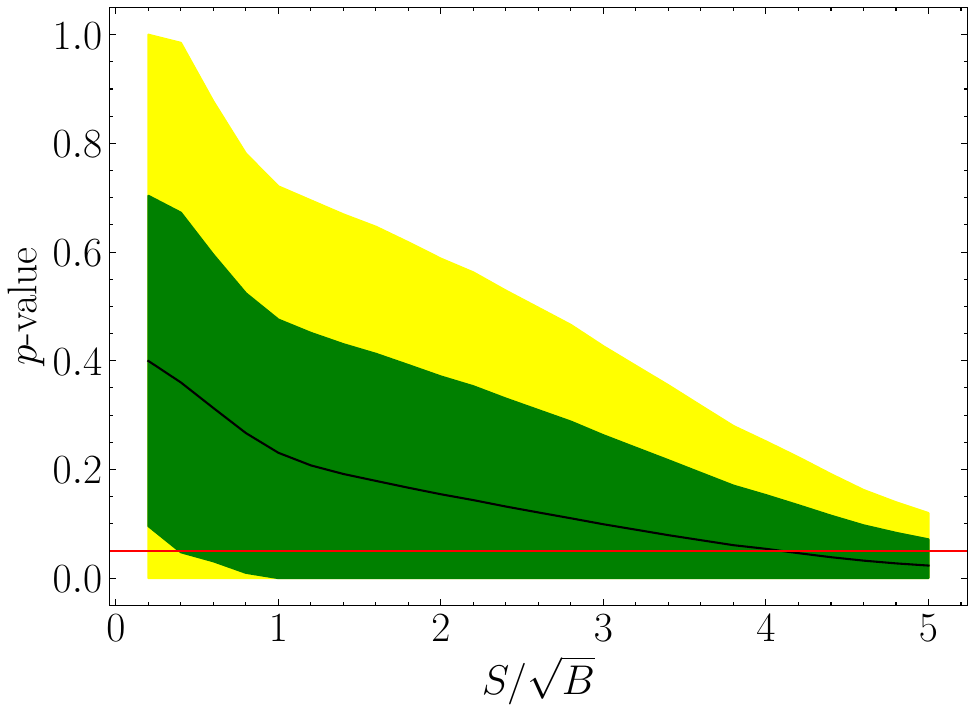}
    \caption{$p$-value as a function of $S/\sqrt{B}$ for the 2HDM signal (left) and the RS signal (right) using Cr as the statistic in the statistical test for the collection of AE with the latent space dimension of $31$. The uncertainty bands correspond to 1 and 2 standard deviations of the $p$-values obtained from the model
    collection, while the red line indicates the $p$-value at $95\%$ confidence level.}
\label{fig:p-values_brazilian_plot}
\end{figure}

As a proof of concept, we start by producing the control and analysis sets from an equal split of the test set.\footnote{In this step, we found that performing a stratified splitting according to the SM processes instead of purely random splitting was crucial to guarantee a stable statistical test. This can be somehow surprising as our dataset has very good statistics across all SM processes. However, random splitting cannot guarantee the correct expected yields of each of the SM processes, which have very different cross-sections and can have a profound impact on the statistical test.} The control sample is composed of background events normalised to the expected yield $B$ at 150~fb$^{-1}$, while the analysis sample is composed of a mixture of background and signal events normalised to $S+B$. We vary $S/\sqrt{B}$ to assess how the Cr permutation test responds to signal proportion in terms of resulting sensitivity. The resulting analysis distribution is then normalised to the total control yield, to base the test on the difference between the distributions' shape and complement a counting experiment.  In~\cref{fig:p-values_brazilian_plot} we show the $p$-values for rejecting the null hypothesis using AE discriminants in the presence of 2HDM and RS signals. Results show no discrimination for 2HDM even with signal contamination up to $S/\sqrt{B} \geq 5$. For the RS signal, the monotonic decrease in $p$-value with the increase of the signal significance is pronounced, contributing to validating our methodology. In this case, we would find evidence to reject $H_{0}$, and therefore evidence supporting the presence of new phenomena, for $S/\sqrt{B} \gtrsim 4$.
Furthermore, the spread of the $p$-values over the collection of AE models is quite broad when there is no sensitivity, as $t_{obs}$ is expected to be distributed similarly to $P(t|H_0)$ under $H_0$, leading to naturally large fluctuations.

\subsection{Results}

In the previous section, we showed the potential of rejecting the SM-only hypothesis in the presence of new phenomena using the permutation test on a semi-supervised discriminant. We now extend the proof of concept to all signals under study, and use it as a metric for testing the dependency of semi-supervised models introduced in~\cref{sec:methods} to the possible untunable hyperparameters, as presented in \cref{sec:hyper} with the ROC AUC.
%-- which the impact on sensitivity as measured by the ROC AUC was detailed in~\cref{sec:hyper} --, and for t
The two test statistics introduced in~\cref{sec:test} are employed, and the signals in the analysis sample are normalised such that $S/\sqrt{B}$=2. This choice poses an interesting scenario wherein the signal significance of a counting experiment is noticeable for a signal expectation but otherwise interpreted as an insignificant fluctuation of the expected background. Therefore, we study whether the statistical test could provide evidence for new phenomena in a regime where it could be easily overlooked, enabling a semi-supervised search to obtain evidence to reject the SM-only hypothesis.

\begin{figure}[]
    \centering
    %MDelta
    \includegraphics[width=\textwidth]{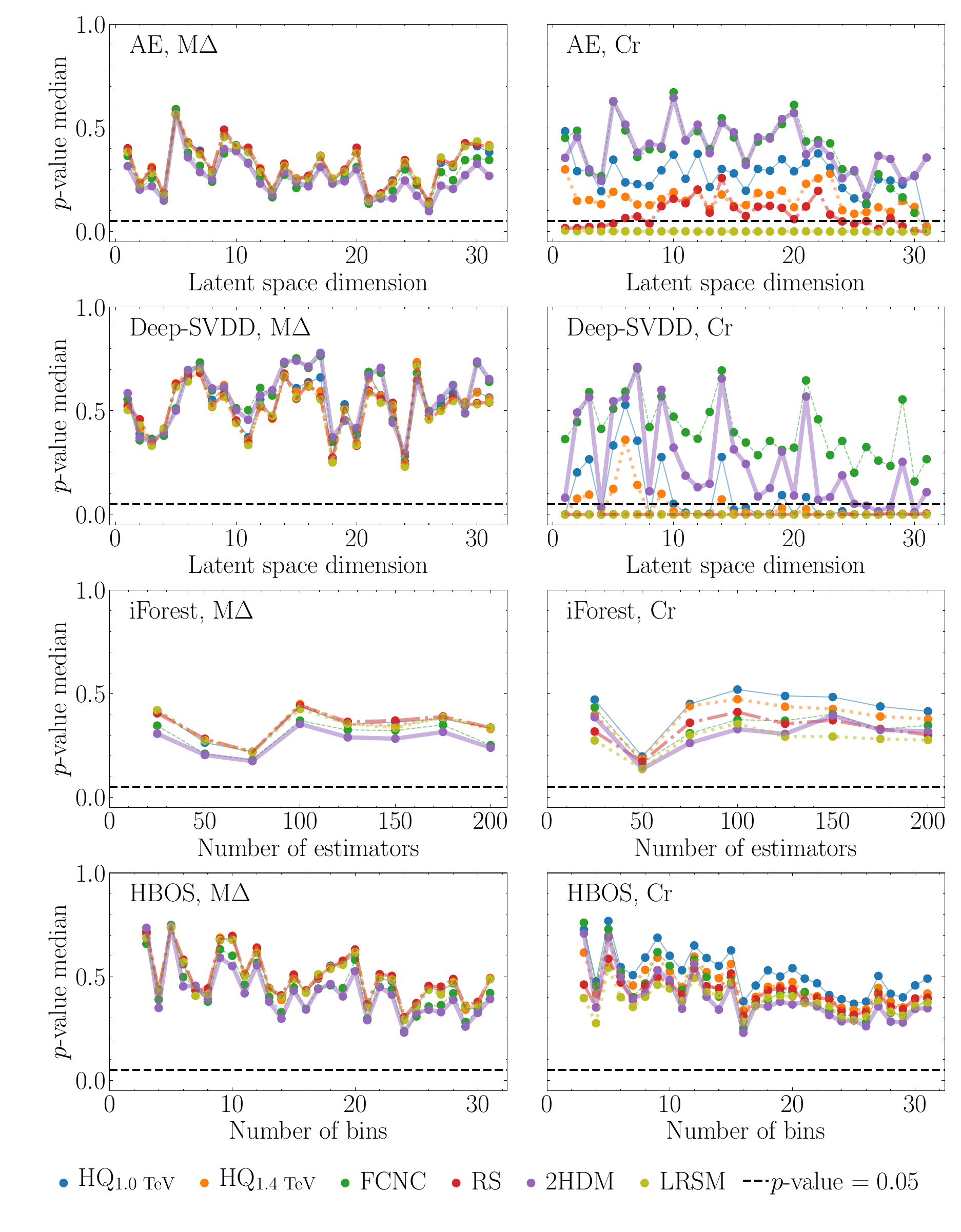} 
    \caption{Median $p$-value for each signal and model collection using M$\Delta$ (left) and Cr (right) as a function of the untunable hyperparameter. The signal in the analysis sample is normalised such as $S/\sqrt{B}$=2.}
    \label{fig:pval_total}
\end{figure}

The full results are presented in~\cref{fig:pval_total} for each signal type using the M$\Delta$ and Cr test statistics, showing the median $p$-value over the collection of $10$ trained models as a function of the untunable hyperparameter.

The first observation standing out is the importance of the choice of test statistics, as M$\Delta$ fails to produce evidence for the presence of new phenomena in $\mathcal{A}$ (median $p$-value is always $>0.05$) across all signals and AD methods irrespective of the value of the untunable hyperparameter. On the other hand, Cr shows to be more promising at capturing evidence for the rejection of the SM-only hypothesis, with sensitivity varying across signals, AD models, and their configurations.

Secondly, the Cr-based results show significantly lower $p$-values for the deep learning methods. This is somehow surprising as the shallow methods consistently outperformed the Deep-SVDD in terms of ROC AUC, c.f.~\cref{fig:roc_auc}. This reinforces the fact that the ROC AUC can be a misleading metric to measure semi-supervised sensitivity, as argued earlier. The increased sensitivity of deep learning can be associated with the long-tailed nature of their outputs. Both shallow methods produce a bounded-value anomaly score: there is a finite number of trees in the iForest and a finite number of bins contributing to the HBOS score. Oppositely, the deep learning models produce unbounded outputs, as an event can be arbitrarily badly reconstructed by the AE or land arbitrarily far from the centre of mass of the distribution in the Deep-SVDD latent space. The combination of these long-tailed distributions with the Cr test statistics, which integrates the differences of the eCDFs across the discriminant domain, explains the increased sensitivity observed. The same effect does not happen when using M$\Delta$, since this test statistic only captures the maximum difference of the eCDFs.

The third important feature is the impact that the untunable hyperparameters have on the sensitivity. 
Overall, the results show no clear dependence on this hyperparameter choice, with no notable trend or an outstanding optimal value. However, the variation with the untunable hyperparameter is still significant indicating that its choice should not be overlooked. The observations suggest that an approach to the problem could be to aggregate results from AD models with different untunable hyperparameters. While in this work we took a visual approach by assessing the $p$-value variation over the different hyperparameter values in plots such as~\cref{fig:pval_total}, other approaches that quantitatively aggregate different $p$-values, such as those presented in~\cite{signalagnostictest_2024}, could provide an alternative way of assessing sensitivity across all values of the hyperparameters.

Finally, the results show some sensitivity complementarity between the AE and the Deep-SVDD, with the Deep-SVDD presenting in general a higher sensitivity to both HQ signals. This further corroborates the hypothesis already pointed out in~\cite{Crispim_Rom_o_2021} that different AD methods capture different notions of anomaly and do not always agree on what anomalous events are.

\section{Conclusions}
\label{sec:conclusions}

In this work, we investigate the influence of untunable hyperparameters of AD methods in the sensitivity of signal-agnostic searches for new physics phenomena. A selection of shallow -- iForest and HBOS -- and deep -- AE and Deep-SVDD -- semi-supervised AD methods are trained and their performance is assessed with a set of BSM signals.

Using the ROC AUC as a metric for signal sensitivity, we conclude that the AE discriminative power is not directly related to the background reconstruction quality. Moreover, no significant variation is observed with the choice of untunable hyperparameters for most BSM signals and across all AD methods. Under this metric, shallow methods outperform the Deep-SVDD for all signals and across all hyperparameter values, while being competitive to the AE, within the uncertainty pertaining to the stochastic nature of the trainings. For all cases except the Deep-SVDD, the ROC AUC was mostly compatible with the one obtained by the most discriminating feature. Since the selection of the best feature is signal-dependent, this proves that the AD methods preserve the discrimination of the best feature in a signal-independent way.

Two test statistics (M$\Delta$ and Cr) were used to assess the sensitivity towards new phenomena where a significance threshold of $p$-value $\leq 0.05$ is associated with evidence for new phenomena, or, more precisely, for the rejection of the SM-only null hypothesis. 
For a signal injection of $S/\sqrt{B} = 2$, our results show that deep methods achieve sensitivity for most signals when using Cr as the test statistic. For M$\Delta$, we see no sensitivity for any method and any of the signals, highlighting the importance of choosing the right test statistic for the discriminant domain. The limitations of the ROC AUC as a measure of signal-agnostic discrimination provide the underlying reasons for the little relation observed between ROC AUC and the $p$-values. However, the same qualitative conclusion is extracted from $p$-values and ROC AUCs considering an unobserved correlation between the signal sensitivity and untunable hyperparameters. Furthermore, the $p$-value metric unveils a significant sensitivity variation across the tested values. We also notice sensitivity complementarity between the AE and the Deep-SVDD, where a combination of the two would, in principle, improve performance. This reinforces the ``no free lunch theorem''~\cite{free_lunch_96}, stating that no single machine learning model can outperform all others for every task, which in the context of AD means that discriminants do not equally agree on what an anomaly is.

Our work also suggests new research avenues to develop the methodology proposed. Namely, the choice of the test statistic is crucial to produce a semi-supervised statistical test of the SM-only hypothesis that is sensitive to new phenomena. Additionally, given how different semi-supervised AD models produce strikingly different sensitivities, it is likely that other AD discriminants would improve upon the ones presented here. Finally, while we only considered a simple permutation test, we leave to future work a more detailed comparison between the permutation test and other approaches to two-sample statistical testing.

In summary, in this work we have shown that the untunable hyperparameter choice in semi-supervised AD models highly impacts the sensitivity of searches for new phenomena, suggesting that strategies built on the aggregation of models should be explored to tackle the problem. Finally, we introduced a method to produce semi-supervised statistical tests on the SM-only null hypothesis, paving the way for purely semi-supervised searches for new phenomena, complementary to supervised searches in experiments.

\section*{Acknowledgments}

This work is supported in part by the Portuguese
Fundação para a Ciência e Tecnologia (FCT) under LA/P/0016/2020 and UID/50007/2023.
FAS is supported by FCT under the research Grant UI/BD/153105/2022. 
MB is supported by FCT under the research Grant SFRH/BD/151006/2021. 
MCR is supported by the STFC under Grant ST/T001011/1.
RP is supported by FCT under Grant 2021.01023.CEECIND.
We would like to acknowledge Centro Nacional de Computação Avançada (CNCA) under 2023.10635.CPCA.A1.

%\newpage

\bibliographystyle{unsrt}
\bibliography{bib}

\end{document}